# Collaborative Intelligent Cross-Camera Video Analytics at Edge: Opportunities and Challenges

Hannaneh Barahouei Pasandi, Tamer Nadeem
Virginia Commonwealth University
{barahoueipash,tnadeem}@vcu.edu

## ABSTRACT

Nowadays, video cameras are deployed in large scale for spatial monitoring of physical places (e.g., surveillance systems in the context of smart cities). The massive camera deployment, however, presents new challenges for analyzing the enormous data, as the cost of high computational overhead of sophisticated deep learning techniques imposes a prohibitive overhead, in terms of energy consumption and processing throughput, on such resource-constrained edge devices. To address these limitations, this paper envisions a collaborative intelligent cross-camera video analytics paradigm at the network edge in which camera nodes adjust their pipelines (e.g., inference) to incorporate correlated observations and shared knowledge from other nodes' contents. By harassing redundant spatio-temporal to reduce the size of the inference search space in one hand, and intelligent collaboration between video nodes on the other, we discuss how such collaborative paradigm can considerably improve accuracy, reduce latency and decrease communication bandwidth compared to non-collaborative baselines. This paper also describes major opportunities and challenges in realizing such a paradigm.

Video analytics plays a key role in smart cities and connected applications such as crowd counting, activity detection, event classification, traffic counting, etc. With recent advances in machine intelligence technologies especially Deep Neural Networks (DNNs), a set of networked cameras enable the use of automated, near real-time analytics for different applications and services. Due to the high demand for computation and storage resources, DNNs are often deployed in the cloud. Therefore, nowadays, video analytics is typically done using a cloud-centric approach where data is passed to a central processor with high computational power. However, this approach introduces several key issues. In particular, executing DNNs inference in the cloud, especially for real-time video analysis, often results in high bandwidth consumption, higher latency, reliability issues, and privacy concerns. Therefore, the high computation and storage requirements of DNNs disrupt their usefulness for local video processing applications in low-cost devices. For example, GoogLeNet model for image classification is just larger than 20 MB and requires about 1.5 billion multiply-add operations per inference per image. Hence, it is infeasible to deploy current DNNs into many devices with low-cost, low-power processors. Worst yet, today video feeds are independently analyzed. Meaning, each camera sends its feed to the cloud individually regardless of considering to share possible valuable information with proximate cameras and spatio-temporal redundancies between the feeds. As a result, the required computation to process the videos can grow significantly.

Motivated by the aforementioned challenges, we envision a new paradigm for collaborative intelligent cross camera video analytics at the edge of the network. We believe that such a paradigm can significantly lower energy consumption, bandwidth overheads, and latency, as well as provide higher accuracy and ensure respecting better privacy by leveraging knowledge sharing and spatio-temporal correlations among cameras. Jian et al. [5] also discuss a vision that leverages cross-camera correlations. However, their main focus is to scale large camera deployment while maintaining sub-linear or constant computation cost grow. Our vision goes beyond leveraging cross-camera correlations, and consider all the aspects of such a paradigm including privacy and adversarial attacks that are not considered in their work.

## 1 CURRENT APPROACHES

Vision algorithm architectures (e.g., CNNs) may consist of millions of parameters to be tuned that require sophisticated computing and storage resources. In this section, we discuss the existing current approaches that bring vision techniques to resource-constrained IoT devices e.g., video nodes.

### 1.1 Downsizing ML Models

One of the approaches to execute the vision tasks on resource-constraint devices is to downsize the model itself. In the following, we categorize the existing approaches and discuss each in brief.

*1.1.1 Model Compression.* In model compression, a dense network is converted to a sparse network that helps to reduce the storage and computational requirements of DNNs. However, this approach is not applicable to all kinds of model architectures. Lane et al. [6] measure different factors that embedded, mobile, and wearable devices can bear for running Deep Learning (DL) algorithms. These factors included measurements of running time, energy consumption, and memory footprint. The study focuses on investigating the behavior of Convolution Neural Networks (CNNs) and DNNs on three hardware platforms that are used in IoT, mobile, and wearable applications.

Model pruning is another technique of compression that focuses on pruning redundant and unnecessary connections and neurons as well as using weight sharing mechanisms. Weight pruning is a widely explored approach to optimize executing CNN models. The key feature of this approach is to select the "appropriate" weights to prune or compress. A recent effort proposed to preferentially prune the weights of nodes that are predicted to be energy-hungry [12].

*1.1.2 Designing Hardware Accelerators.* Designing specific hardware and circuits is another ongoing research direction aiming to optimize the energy efficiency, and memory footprint of the models in IoT devices. The focus of such research works is on the inference time of Deep Learning (DL) models. As an example of such work, DeepEye [8] is a distinctive wearable that is capable of executing cloud-scale deep vision models entirely on a single device without offloading to the cloud. Two core enablers of DeepEye are the following. The first is the hardware design that is powered by a Qualcomm Snapdragon 410 processor and a custom integrated carrier board consisting of a 5 megapixel camera sensor. The second is an inference pipeline specifically optimized to cope with the needs of multiple models.

## 1.2 Enhancing Edge Resources

Deploying shared resources at the Internet edge and pushing computation close to the data sources can benefit many applications requiring low latency and high privacy. Liu et al. propose EdgeEye - an edge service framework for real-time intelligent video analytics [7]. EdgeEye uses a powerful edge server in which different devices can offload their computation to it. Current research approaches focus more on training than inference. The inference performance will not be optimal if we directly use DL frameworks to execute a DNN. Chip vendors like Intel and Nvidia provide optimized inference engines for CPUs and GPUs, which EdgeEye leverages these implementations to realize highly efficient inference services.

## 1.3 Workload offloading

Another direction of the recent efforts focuses on developing schemes for deciding where to run ML models; locally on the device or on a server (edge or cloud). CoINF [11] describes a framework in which a model can be split into sub-models using model partitioning technique to be executed on wearable devices and phones. Their results show that for some scenarios, partial offloading of he execution of a model to a phone outperforms the binary decision. This is because an internal layer inside the model may yield a small intermediate output compared to the original input size, and thus reduces the data-transmission delay. Authors in [3] implement a DL inference offloading system for robotic vehicles, which takes advantage of a hardware accelerator from Intel and a partitioning approach to execute DL models on the edge for these vehicles.

## 2 OPPORTUNITIES

The main metrics of interest in collaborative cross-camera video analytics applications at the edge include training phase, inference accuracy, processing cost, network bandwidth, and near real-time analysis. In the following, we discuss the opportunities that our novel collaborative intelligent paradigm could bring to each of the aforementioned metrics of interest.

## 2.1 Smarter Models

Different cameras could be deployed in different zones in which each zone could be managed by a different organization, observe unique behaviors, and have different constraints. Under such conditions, each camera would be limited to be trained using its own local data for better accuracy and privacy. For example, human behaviors in a camera facing a parking lot would be completely different than the ones in a camera facing a nightclub, and hence it would be more accurate to build a customized model for each camera based on its local data. Another example when cameras are deployed in sensitive areas in which each will be limited to its local data since they can't share data with others for privacy concerns. Yet, sharing certain model knowledge between specific cameras would improve the accuracy of these specific models. In doing this, it is important to identify which camera(s) to learn from and what knowledge to transfer while maintaining a low overhead. Such paradigm of allowing cameras to be trained using their local data and then sharing knowledge of the built models between relevant cameras only will enable us to have smarter customized models, lower latency, less communication overhead, less power consumption, all while ensuring privacy.

## 2.2 Resource Saving

In cross-camera applications, there are often far fewer objects of interest than cameras. In addition, with dense deployment, adjacent cameras will have significant overlap between their views. Assuming a scenario with two spatially correlated cameras namely *Camera A* and *Camera B* that have overlapping views, we are confident that the overlapped Field of View (FoV) should be analyzed once. In this scenario, we assume that both cameras execute the same vision task e.g., object detection. When both cameras start the inference step, at early stages of the inference, they can share execution states of their models. By using techniques such as feature matching, they can realize whether both are detecting the same object. If so, one of the cameras (e.g., *Camera B*) could stop its inference execution saving its resources. In addition, *Camera A* could possibly offload parts of its model execution to *Camera B* speeding-up the object detection process. In such an approach, we need to consider the trade-off between transmission time overhead and inference execution time. Another possible approach is to use a Siamese Neural Network [1] on input frames of both cameras before inference step in order to measure the similarities between frames' extracted features. However, these approaches requires cameras to be synchronized.

## 2.3 Enhancing Accuracy

**Sharing camera input feeds across models.** Assuming a scenario where several surveillance cameras are installed in various locations in which some of these locations could negatively impact some of the video analysis tasks. For example, a camera located at an elevated level (e.g. on top of a pole) could suffer from a low accuracy of facial recognition. This issue could be addressed by retraining the model using the input feed of another camera node that has an eye-level view for example. Cameras that are installed in vulnerable positions could suffer of low quality images at different times of the day or under various weather condition (e.g., under extreme luminance during the day or a rainy weather condition), which results in lower accuracy due to poor quality of input data. However, other cameras installed in a better position that have an overlapping FoVs would have better inference performance. Therefore, such cameras can complement each other via collaboratively sharing their inputs.

**Collaborative inference by sharing extracted features across cameras.** Using an ensemble of identical models to make a prediction is an accepted approach for enhancing inference accuracy [4]. This technique could also be applied to model ensembles consisting of multiple camera video pipelines that have a different view of the same object. Let's assume the following scenario to show the potential of boosting inference performance by sharing features. Our scenario is motivated by Jian et al. [5] observation on "DukeMTMC" dataset in which objects that appear in one camera are going to appear in the nearby cameras about 90% of the time. Therefore, detecting spatio-temporal correlations between cameras can be useful in terms of reducing execution time and computation resources by sharing the learned knowledge and removing redundancy across cameras. In this scenario, for example, *Camera A* detects an object first and then performs object detection task. In between or after the inference is done, *Camera A* shares the extracted features with *Camera B* that is in the proximity of *Camera A*. When *Camera B* detects an object, it performs feature matching to check whether it has detected the same object as *Camera A*. If features match, then *Camera B* does not need to execute the complete object detection inference. Moreover, by combining object detection results of both cameras, even when matching features, higher confidence in terms of result accuracy of both the two cameras could be achieved.

## 2.4 Near Real-time Analytics

Given the dense deployment of camera nodes in which several would be idle or underutilized, it is desired to utilize these idle resources in speeding up the individual cameras in order to have near real-time video analytics. One popular approach is to split the input frame into subframes and distribute these subframes to adjacent idle nodes for parallel processing to improve the inference time.

In current approaches, different camera nodes use an identical model for a specific task. We argue that we could get near real-time video analytics by redesigning current models as co-models, which are instants of the original model but work on multiple devices collaboratively. As an illustrative example, Figure 1 shows two instances of the same decision tree that tries to identify the range of a real number. Each tree will require a maximum of nine comparisons to identify the range. However, if both decision trees were executed on different nodes in which they are able to exchange information with each other after each comparison, the maximum number of comparisons will be reduced to five comparisons only that approximately cut the execution time to the half. We believe that co-model design is an interesting research direction that will attract the attention of the research community.

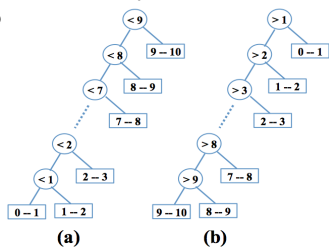

**Figure 1: Instances of the same decision tree**

## 2.5 Improved Privacy

The use of computer vision technologies is not limited to the rapid adoption of facial recognition technologies but is also extended to facial expression recognition, scene recognition, etc. These developments raise privacy concerns regarding the collection and the use of sensitive personal data. These concerns can grow to the extent that regulators and authorities take serious actions with regards to these technologies. As an example, recently, San Francisco banned facial recognition technology [2]. Most of the current privacy-aware video streaming approaches involve denaturing, which means a content-based modification of images or video frames guided by a privacy policy. For example, a mechanism in which objects in images that have been either physically or virtually tagged by users are blocked [10].

In our paradigm, privacy can be achieved by defining policies in which cameras that capture sensitive information will perform locally certain stages of the inference pipeline that contain those sensitive data. Although some of the information might be kept locally, enabling these cameras to work collaboratively will help their local observations to enhance the overall accuracy of the desired model. In the following, we describe a scenario about how privacy can be achieved by using a shared model among cameras. In this scenario, cameras use the same shared model for a specific vision task which is trained by non-sensitive frames of each camera. *Camera x* performs inference locally using this model for every new sample. The new sensitive data captured by *Camera x* are then used to locally retrain and obtain a customized instance of the same model for *Camera x* without sharing its sensitive data. To assure accurate inferences, *Camera x* could also validate its accuracy by cross-verifying with proximate cameras that are performing the same vision task.

## 3 CHALLENGES

### 3.1 Handling Dynamic Intermittent Devices and FoVs

When camera nodes are stationary, the spatio-temporal correlations are relatively invariant which implies a stable collaboration among cameras. However, sometimes the camera nodes (e.g., phone, wearable, drones, etc.) are mobile, and most likely their relative locations and FoV changes dynamically. This requires sophisticated techniques to continuously update collaborative inference based on their dynamic movements and determine strategies with which node to collaborate. Even in the case of static cameras, sometimes the camera's FoV changes. For instance, a camera that faces a school, and then later the school is rebuilt to be a parking lot. In such a scenario, the camera requires a new training phase to adapt its inference to such change. In such scenarios, having online training could also play a role.

### 3.2 Detecting Adversarial Devices

The performance and accuracy of our paradigm could be affected significantly with the presence of adversarial cameras. Since all nodes including adversarial cameras share their inference with the proximity nodes, we need mechanisms to first detect and isolate such adversarial nodes. Authors in [9], describe their early efforts in enabling resilient-collaboration, via the use of reputation scores. This mechanism is, however, based on the assumption that a malicious camera node continuously and randomly modifies its shared information features (the bounding boxes of detected people objects). However, malicious cameras may expose more complicated and sinister behavior e.g., failing to report accurate features only for selected individuals of high interest. Moreover, their proposed histogram-based approach is more applicable only for those cameras with concurrent spatial and temporal overlap. Therefore, our paradigm requires more sophisticated strategies to identify malicious activities and handle a wide variety of adversarial behavior.

### 3.3 Dealing with Small Training Datasets

Sometimes a camera is installed in a premise such that it does not have many useful samples to learn from. For instance, one camera is placed in a main hallway and detect many candidate samples, while another camera detects very few at the same time. One approach to overcome this challenge could be to share the samples of the camera in the hallway with the other camera to train its model without sacrificing privacy. In addition, as stated in [5], such scenarios with heterogeneous workload could also provide an opportunity for dynamic offloading and utilizing idle resources of less-utilized cameras.

### 3.4 Challenges in Training and Inference Phase

In the following, we describe a training phase challenge in our paradigm through an experiment of person Re-Identification (Re-Id) task on MSMT-V2 [13] dataset that includes 4,101 identities and 15 cameras in the network.

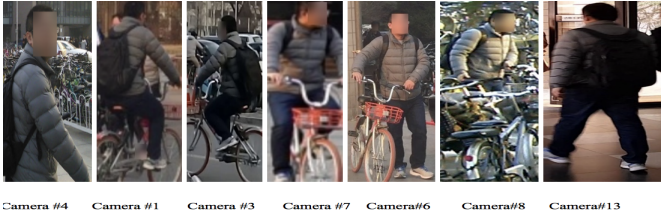

Figure 2: Sample Frames of The Same Identity Captured Across Different Cameras.

In this experiment, we assume we know the cameras with overlapping FoVs beforehand. We use the feeds from *Camera x* to train its model. Then, we extract a series of useful features and share them with *Camera y*. Based on the topology of the networked cameras, we identify the input feeds of cameras with overlapping (in this case, cameras #4,6,7) and other cameras to show that the potential cameras with overlapping views and insignificant luminance conditions can actually contribute to a higher score in person Re-ID task.

As shown in Table 1 when cameras namely 4 and 7 share their input feeds with Camera 6, the score improves by almost %13. On the other hand, most importantly when the majority of the cameras share their input with Camera 6, the score drops to %27. This drop is justifiable because some of the difference in time, weather condition, angles that these feeds are captured. As an illustrative example, Fig. 2 shows sample frames of the same identity captured across different cameras. Therefore, in this collaborative paradigm, the challenge is to identify whom to collaborate with, and what to learn from other nodes. Similar to what discussed in [5, 9], this challenge requires discovering correlations between the nodes. The collaborative cross-camera concept requires camera nodes to autonomously discovers spatio-temporal correlations among nearby cameras to enables collaboration among themselves. To discover such correlations, one approach could be to learn the correlations from the inference results. For instance, if two cameras identify the same object in a short time window, they potentially have a content correlation.

| Camera feed used | Score |
|---|---|
| Camera #6 | 0.37547 |
| Camera #4,6,7 | 0.50544 |
| Camera #1,3,4,6, 7,8,11,12,13 | 0.27365 |

Table 1: Comparing scores for person Re-ID task using different camera feeds.

## 4 CONCLUSION

This paper describes our vision of a collaborative intelligent cross-camera video analytics at the edge, a paradigm where video nodes are used to enable collaborative execution of machine intelligence tasks on resource-constrained cameras on the network edge. We believe that such intelligent cross-camera collaboration can significantly lower energy, bandwidth overheads and latency, and provide better accuracy and ensures better privacy. We have highlighted the new opportunities and key challenges associated with realizing such a paradigm. We also motivated the need for such a paradigm through some examples. Although we only focused on collaborative cross-camera video analytics application, we believe such a paradigm could also be extended to collaboration between other types of IoT nodes/sensors such as audio, motion, etc.